 \documentclass[aps,pre,twocolumn,float]{revtex4}

 \usepackage{graphicx}
 \usepackage{amsmath,bm,epsfig}
\def \ed {\end{document}}
\def\Fbox#1{\vskip1ex\hbox to 8.5cm{\hfil\fboxsep0.3cm\fbox{%
  \parbox{8.0cm}{#1}}\hfil}\vskip1ex\noindent}  %%  {TEXT} in BOX

\newcommand{\eq}[1]{(\ref{#1})}%%  requires \eq{label}
\newcommand{\Eq}[1]{Eq.~(\ref{#1})}%%  requires \eq{label}
\newcommand{\Eqs}[1]{Eqs.~(\ref{#1})}%%  requires \eq{label}
\newcommand{\Fig}[1]{Fig.~\ref{#1}}%%  requires \Fef{label}
\newcommand{\Figs}[1]{Figs.~\ref{#1}}%%  requires \Fef{label}
\newcommand{\Sec}[1]{Sec.~\ref{#1}}%%  requires \Fef{label}
\newcommand{\Ref}[1]{Ref.~\cite{#1}}%%  requires \Fef{label}
\def\be{\begin{equation}}\def\ee{\end{equation}}
\def\bea{\begin{eqnarray}}\def\eea{\end{eqnarray}}
\def\bse{\begin{subequations}}\def\ese{\end{subequations}}
\newcommand{\BE}[1]{\begin{equation}\label{#1}}
\newcommand{\BEA}[1]{\begin{eqnarray}\label{#1}}
\newcommand{\BSE}[1]{\begin{subequations}\label{#1}}

\let \= \equiv \let\*\cdot \let\~\widetilde \let\^\widehat \let\-\overline
\let\p\partial

  \def\1{\bm1} 

\def\<{\left\langle}    \def\>{\right\rangle}
\def\({\left(}          \def\){\right)}
 \def \[ {\left [} \def \] {\right ]}

\renewcommand{\d}{\delta}
\newcommand{\e}{\epsilon}\newcommand{\ve}{\varepsilon}
\renewcommand{\o}{\omega} 
\renewcommand{\L}{\Lambda}

\def\k{\kappa}

\newcommand{\B}[1]{{\bm{#1}}}%% Bold Roman & Greek Lower & Upper Case
\newcommand{\C}[1]{{\mathcal{#1}}}    %%   Calligrapfic Upper case
\renewcommand{\sb}[1]{_{\text {#1}}}  %% sub-   for lower case
\newcommand{\Sp}[1]{^{^{\text {#1}}}} %% Super- for Upper case
\def\Sb#1{_{\scriptscriptstyle\rm{#1}}}

\begin{document}

\title{ Gradual eddy-wave crossover in superfluid turbulence.  }
\author{Victor S. L'vov$^{*\, \ddag}$,      Sergey V. Nazarenko$^\dag$
and Oleksii Rudenko$^*$}
  \affiliation{$^*$Department of Chemical
Physics, The Weizmann Institute of Science, Rehovot 76100, Israel}

\affiliation{ $^\dag$University of Warwick, Mathematics Institute,
Coventry, CV4 7AL, UK}

\affiliation{ $^\ddag$ Department  of Theoretical Physics, Institute for Magnetism, Ukraine National Ac. of Sci., Kiev, Ukraine}
\begin{abstract}
We revise the theory of superfluid turbulence near the absolute zero of temperature
and suggest a model with differential approximation for the energy fluxes in the $k$-space, $\ve\Sb{HD}(k)$ and  $\ve\Sb{KW}(k)$, carried, respectively, by the
collective hydrodynamic (HD) motions of quantized vortex lines and by their individual uncorrelated motions  known as Kelvin waves (KW).  The model predicts energy spectra of the HD and the KW components of the system,  $\C E\Sb{HD}(k)$ and $\C E\Sb{KW}(k)$, which experience a smooth crossover between different regimes of motion over a finite range of scales. For an experimentally relevant  range of $\L\equiv \ln (\ell/a)$ ($\ell$ is the mean intervortex separation and $a$ is the vortex core radius) between 10 and 30 the total energy flux $\ve=\ve\Sb{HD}(k)+\ve\Sb{KW}(k)$ and the total energy spectrum $\C E(k)=\C E\Sb{HD}(k)+\C E\Sb{KW}(k)$ are dominated by the HD motions for $k  \lesssim 2/\ell$. In this region $\C E(k)$ follows the HD spectrum with constant energy flux $\ve\simeq \ve\Sb{HD}=$const.:  $\C E(k)\propto k^{-5/3}$ for smaller $k$ and tends   to equipartition of  the HD energy  $\C E(k)\propto k^2$  for larger $k$.  This bottleneck accumulation of the energy spectrum  is milder than the one predicted before in \cite{lnr} based on a model with sharp HD-KW transition. For $\L=30$, it results in  prediction for the effective viscosity  $\nu^{\,\prime} \simeq 0.002 \kappa$ ($\kappa$ is the circulation quantum) which is in a reasonable agreement with its experimental value in $^4$He low-temperature experiment $\approx 0.003 \kappa$.  For $k \gtrsim 2/\ell$, the energy spectrum is dominated by the KW component:
%, however the main contribution to the energy flux still goes from the HD turbulence.
 almost flux-less KW component close to the  thermodynamic equilibrium, $\C E\approx \C E\Sb{KW}\approx $const
at smaller $k$
%  (while HD energy exponentially decays, $\C E\Sb{HD}\propto \exp(- k\ell)$). Only at relatively large $k>k\sb{fl}\approx$ 24 (for $\L\sim 10 - 30$) energy flux is fully supported by
and the KW cascade spectrum $\C E(k)\to \C E\Sb{KW}(k)\propto k^{-7/5}$ at larger $k$.
\end{abstract}

\maketitle
\section{Introduction}

Liquid $^4$He and $^3$He at very low temperatures can be viewed as a perfect superfluid without any normal fluid present. Turbulence is a very common state for such superfluids, and it comprises one of the most interesting subjects in physics with exciting recent developments in turbulence;  see, e.g.~\cite{VinenNiemela,Vinen2008,Tsubota2008,rev4,exp0,araki,stalp,exp,exp1,Vin03-PRL,cn,exp2,exp3,exp4,Helsinki-exp,RocheBarenghi}. Superfluid turbulence consists of a tangle of quantized vortex lines~\cite{VinenNiemela,Vinen2008,Tsubota2008,rev4,1,2}. How is superfluid turbulence related to usual hydrodynamical turbulence? On the one hand, at the scales greater than  the mean distance between the inter-vortex separation distance one can expect the vortex discreteness to be unimportant and, therefore, superfluid and hydrodynamical turbulence should have similar properties at these scales. This can be true, of course, if the superfluid vortex tangle is not completely random but polarized and organized into vortex bundles which at large scales form similar motions as would continuous hydrodynamic eddies. In turn, such a vortex polarization can be either introduced by the external forcing (yet to be understood how), or it can occur due to a (yet to be found) self-organization mechanism. On the other hand, since there is no viscosity, the energy would cascade in superfluid turbulence downscale without loss until it reaches to the small scales where the quantum discreteness of vorticity is important. It is believed that at this point the Kolmogorov-type (K41) eddy dominated cascade is replaced by a cascade due to nonlinearly interacting Kelvin waves. Kelvin wave cascade takes energy further downscale where it can be radiated away by phonons.

Although the overall picture of superfluid turbulence described above seems quite reasonable, some important details of this picture are yet to be established. A particularly interesting question is about the structure of the crossover between the eddy dominated and the wave dominated regions of the spectrum. As it was pointed out in our recent paper~\cite{lnr}, the nonlinear transfer mechanisms among weakly nonlinear Kelvin waves on discrete vortex lines is less efficient than the energy transfers due to the strongly nonlinear eddy-eddy interactions in continuous fluids. This results is an energy cascade stagnation  at the crossover from the collective eddy dominated to the single-vortex wave dominated scales. The main message of paper \cite{lnr} is that such a \emph{bottleneck} phenomenon is robust and common for all the situations where the energy cascades experiences a continuous-to-discrete transition, and the details of particular mechanism of this transition are secondary. Indeed, most discrete physical processes are less efficient than their continuous  counterparts~\footnote{It is interesting to make comparison with turbulence of weakly nonlinear waves where the main energy transfer mechanism is due to wavenumber and frequency resonances. In bounded volumes the set of wave modes is discrete and there are much less resonances between them than in the continuous case, so the energy cascades between scales are significantly suppressed.}. On the other hand, particular mechanisms of the continuous-to-discrete transition can obviously lead to different strengths of the bottleneck.  The quantitative measure of the bottleneck is the value of rms vorticity, because this is the quantity which enters into the definition of the effective viscosity which is experimentally observable via measuring decay of the vortex line density \cite{stalp,VinenNiemela} (see below).

Paper \cite{lnr} considers  the  bottleneck mechanism under the simplest assumption that a sharp transition from the K41 eddy dominated cascade
to the Kelvin wave weak turbulence occurs at the mean inter-vortex
separation scale $\ell$. In this case, the vortex line reconnections
provide a mechanism to transfer the energy from the eddy to the wave
motions, but their role for the energy cascade itself
was neglected (even though it was noted that the bottleneck strength can
be affected if this role was taken into account).
On the other hand, paper \cite{KS07} considered another extreme
when the reconnections are the key process for the crossover cascade which
was suggested to go through three different stages in a rather narrow range
of scales of width $\Lambda$, -- bundle-dominated,  nearest neighbor and
 self reconnections. In spite of this rather unrealistic construction,
 the end result was still a bottleneck and reduction of the effective
 viscosity, though by a smaller factor than predicted in \cite{lnr},
 $\Lambda$ instead of $\Lambda^5$.

 In the present paper we   neglect the role of
 reconnections for the cascade process because, as we argued in  \cite{lnr},
 the reconnections are strongly inhibited within the polarized  vortex bundles,
 and their occurrence is limited to the edges of these bundles.
 Since the volume in between of the bundles is small compared to the
 volume inside of these bundles, it seems natural to assume that the main
 contribution to the cascade will be due to nonlinear dynamics of non-reconnecting
 vortex lines inside the vortex bundles, even though it is still possible that the
 reconnections can adjust the strength of the bottleneck, particularly if the
 K41 range is not too large and the turbulence polarization is reduced.
 The final answer about the role of reconnections should, of course, be sought in
 experimental and numerical data.
 In the present paper, we extend the analysis of  \cite{lnr} by taking into account
 the fact that Kelvin waves can be generated and play a role in the energy cascade
 at the scales greater than $\ell$, and that the transition from the eddy to the wave
 motions occurs over an extended range of scales rather than sharply.
 %However, let us start by a brief description of the bottleneck effect suggested in \cite{lnr}
 %based on the assumption of a sharp crossover at the scale $\ell$.

\section{\label{s:prel} Bottleneck  scenario with   sharp crossovers }

\subsection{\label{ss:dim}Dimensional- and velocity-crossover scales}

When the "bottleneck" effect was first described in~\cite{lnr}, it was
assumed that the crossover from the eddy motions to the wave motions happens
sharply at the scale $\simeq \ell$, i.e. mean intervortex distance, or in the $k$-space at $k\simeq k\sb{dim}$, where
\BSE{X}
\BE{XA}
   k\sb{dim} \ell\simeq 1\ .
 \ee
 Subscript ``~$\sb{dim}$~" reminds that  estimate~\eq{XA} follows from
the simplest possible dimensional reasoning. Besides, it  roughly corresponds to an idea that
at the scales larger than $\ell$ the vortex lines must be polarized and
form bundles which would correspond, in a course grained sense, usual hydrodynamic
eddies, while for $k>k\sb{dim}$ the vortex motions can be viewed as oscillations independently
happening on individual vortex lines, i.e. as 1D Kelvin waves.
On the other hand, it was also remarked in \cite{lnr} that the
self-induced motion of the vortex line can get faster than its motion due
to the collective interaction with the other vortices in the bundle
already at the scales  $k \gtrsim k\sb{vel}$, where
\BE{XB}
k\sb{vel}\ell \simeq   \sqrt {2\big / \Lambda}\ .
\ee%%
Subscript ``~$\sb{vel}$~" reminds that  estimate~\eq{XB} follows from comparison of the self-induced velocity with cross-velocity induced at a given vortex line by nearby $\ell$-distant vortex line. A more detailed discussion of this characteristic scale, which clarifies factor ``~2~" under the square root,  is given below in Appendix \ref{sss:prel}.
Thus, Kelvin waves can be expected to be present in some form already  in the wave-vector range
  from $k\sb{vel}$ to $k\sb{dim}$ where they would
coexist with the collective/eddy motions.
It was argued, however, that due to their oscillatory character
the waves would contribute much less into the cumulative motion of the vortex
bundles in this scale range.

Before going into details what occurs in the transition range
  \BE{ks}
     k\sb{vel}  \lesssim k\lesssim k\sb{dim} \,,
 \ee\ese%%
  it is worthwhile to reconsider some aspects of the problem under the simplest assumption that only one crossover scale is relevant. This is the subject of the current section.

\subsection{\label{ss:sharp}Bottleneck predictions in the presence of dimensional- and velocity-crossovers}
Here we reconsider the simplest scenarios of the eddy-wave transition with  a sharp crossover with the only difference from~\cite{lnr} that the crossover scale  $k_*$  is not necessarily at $k\sb{dim} \simeq 1/\ell$ but lies somewhere  in the range~\eq{ks}. To this end, we remind the Kozik-Svistunov~\cite{KS04} spectrum of Kelvin waves, in the form suggested in~\cite{lnr}:
\BE{KS} %%
\C E\Sb{KW}(k)\simeq  \Lambda
  \big(  \k^7 \ve \big  /\ell^{\,8} \big)^{1\!/5}\, |k|^{-7/5} .
\ee%%
Here $\C E\Sb{KW}(k)$ is the one-dimensional (in the $\B k$-space)    energy density of Kelvin waves, normalized such that  $E\Sb{KW}=\int \C E\Sb{KW}(k)\, d k$ is their total energy in unite volume, $\kappa$ is the quantum circulation and $\ve$ is the energy flux over scales.
Parameters $\ve$ and $\ell$  are mutually dependent, and their relation follows from the
expression for the rms  vorticity in the system of quantum
filaments, $\sqrt{\<|\o|^2\>} \simeq \k \ell^{-2}$. The later  can be found for well developed turbulence when  $\<|\o|^2\>$  is dominated by
the classical-quantum crossover scale $k_*$. In the present case
\BE{est-our}%%
\<|\o|\>^2   = 2 \int k^2 \C E\Sb{HD}(k) \, dk
\simeq  k^3_* \, \C E\Sb{KW}(k_*)\,,
\ee
which, after substitution of (\ref{KS}), gives
 \BE{eps}
 \ve \simeq {\k^3 \big /  \Lambda^5 \ell^{12}k_*^8  }\ .%%
\ee%%
Factor 2 in \Eq{est-our} follows from summation over vector indexes under assumption of isotropy of turbulent spectra.  Equation~\eq{eps} corresponds to the effective viscosity
\BE{nu-p}
 \nu^{\,\prime} = { \ve \ell^4 \over \k^2 } \simeq  \k \big /  \Lambda^{5}(k_* \ell)^8\ .%%
\ee
Under the simplest assumption $k_*\simeq k\sb{dim} \simeq 1/\ell$ one gets the value, reported in~\cite{lnr}: $\nu^{\,\prime}\simeq \k/\Lambda^5$.  For the sharp crossover at the velocity-crossover scale $k\sb{vel}\simeq 1/ (\ell \sqrt{\Lambda})$, where the self-induced velocity is of the order of the   cross-induced velocity  (see below), one gets $\nu^{\,\prime}\simeq \k/\Lambda$. As we argued we expect that the true value of the crossover scale is somewhere in the  region~\eq{ks}. Thus   one   expects that the true value of the bottleneck and corresponding
$\nu^{\,\prime}$ is somewhere in between of the two extreme values
\BE{nu_range}
 \k \Lambda^{-5} \lesssim \nu^{\,\prime}  \lesssim  \k \Lambda^{-1}.%%
\ee
Note that formally the value $\nu^{\,\prime} \simeq  \k \Lambda^{-1}$
is the same as the one predicted by the Kozik and Svistunov approach based on reconnections
\cite{KS07}.

\subsection{\label{ss:X} Bottleneck at sharp amplitude-crossover}
Let us now check the consistency of the assumed in the previous section sharp
crossover at some $k_*$ in the interval~\eq{ks}. To this goal let us evaluate the amplitude $h(k_*)$ of the Kelvin
waves  at this scale. Obviously, for consistency this amplitude   must remain
less than the intervortex separation $h(k_*)\lesssim \ell$. This allows one to introduce the \emph{amplitude-crossover scale} $k\sb{amp}$, at which
\BSE{S1} \BE{check}
h(k\sb{amp}) \simeq  \ell\ .%%
\ee%%
The estimate for $h(k)$ can be obtained from the Hamiltonian of Kelvin wave in the so-called local-induction approximation (see, e.g., Eq.~(5b) in our Ref.~\cite{lnr}):
\BE{h}
h(k)\simeq  \sqrt{  \C E\Sb{KW}(k) \big / \Lambda \k^2 k^3}.
\ee
Substituting here $\C E\Sb{KW}(k)$ from \Eq{KS} and using \Eqs{eps} and \eq{check} one gets estimate
\BE{Xamp}
k\sb{amp}\ell \simeq  \sqrt[6]{2\big / \Lambda}\,,
\ee\ese%%
which is inside of the region~\eq{nu_range}. The subscript ``$\sb{amp}$" reminds that estimate~\eq{Xamp} follows from comparison of   the wave amplitude with the intervortex distance. Factor ``~2~" under the root is put by analogy with \Eq{XB} to ensure that $k\sb{vel}< k\sb{amp}$ for any $\L$. Assuming a sharp crossover at this scale one gets from \Eq{nu-p}:
\BE{nu-prime}
 \nu^{\,\prime} \simeq  \k \Lambda^{-11/3}\,, %%
\ee
which is, as expected, within the range~\eq{nu_range}.

So, in order the waves amplitude to be less than the intervortex distance, the inequality $k > k\sb{amp}$ must hold.
%The problem with estimates~\eq{XB} and \eq{S1}
The problem is to clarify what is going on in the interval%%
\BE{int}
k\sb{vel}\lesssim k \lesssim  k\sb{amp}\,,
 \ee
 where formally computed [with the Kelvin-wave spectrum~\eq{KS}] wave amplitude $h$ exceeds the intervortex distance $\ell$, which cannot physically happen.   For example, at the scale $k\sb{vel}$ one gets from \Eq{h}:
 $h\simeq \Lambda \ell \gg \ell$.
   On the other hand, the motions with $k> k\sb {vel}$ cannot be considered as pure collective, because the cross-velocity (which is the influence of the motion of one vortex line in the place of another one) is smaller than the self-induced velocity (for more detailed discussion of this question, see Appendix~\ref{sss:prel}).

 Our scenario, is that in the interval~\eq{int} the growth
of the wave amplitude on a particular vortex line
would be arrested by the adjacent vortex lines
in the bundle which would "get in the way".
Speculations of similar type of  Kozik and Svistunov
\cite{KS07} lead them to a suggestion that the
hydrodynamic and the wave turbulence ranges are separated
by the range of scales where the energy cascade is dominated by
the vortex line reconnections.
On the other hand, it was pointed out in \cite{lnr} that,
because the vortex lines in turbulence must be polarized and
organizes in bundles, the reconnection process must be suppressed
and pushed to small volumes in between of the vortex bundles.
Instead of a reconnection, one can expect a restriction of the
wave motion of an individual vortex line when it grows in
amplitude and tries to push
close to  the other vortex lines in the bundle.
Naturally the growth of such a wave would get arrested
at the amplitude when the inter-vortex energy (which grows due
to shortening of the distance to the considered vortex
line) becomes equal to the vortex self-energy.

This leads us to the following physical model of
turbulence in the range~\eq{int}.
In the $\textbf{x}$-space, turbulence consists of
vortex bundles with  a fractal
structure.
Each vortex bundle which is made of denser
sub-bundles, such that the mean separation of lines within
the sub-bundle is $\ll \ell$ and the mean distance between the
sub-bundles is $\gg \ell$. In turn, each sub-bundle consists of
even denser sub-sub-bundles, etc. The density of vortex lines
within a particular sub-bundle is such that at the scale
 of this sub-bundle the self-energy [which can be considered as the energy of the Kelvin waves $\C E\Sb{KW}(k)$] and inter-vortex
energies [which should be associated with the hydrodynamic energy $\C E\Sb{HD}(k)$]  are balanced. This corresponds to condition in
the $\textbf{k}$-space,
\BE{bal}
\C E\Sb{HD}(k)  \simeq \C E\Sb{KW}(k)\,,
\ee
in the range~\eq{int}.   For $k>k\sb{amp}$, Kelvin waves can propagate
on an individual vortex line without approaching to (and being influenced
by) the adjacent vortex lines. In the other words, the range
 $k>k\sb{amp}$ is   dominated by the Kelvin wave turbulence and
 the role of the eddy component will  be clarified below.

As we see, our corrected  scenario which takes into account
that the eddy/wave crossover occurs over a finite range of scales
predicts a bottleneck value which is in between of the
values obtained by assuming sharp transitions at the
scales $\ell$ and $\ell \sqrt \Lambda$, respectively.

\section{\label{s:dif} Finite crossover range model}
The goal of this section is to relax the
simplified assumption that the bottleneck happens at some sharp crossover scale and to describe in a simple manner
the transition regimes around the characteristic scales,  introduced in the previous Section.
The first step in this direction is to revise the
differential approximation for the cascades of turbulent energy; this is done in the following Subsection.

\subsection{\label{ss:dif} Differential approximation for the cascades of turbulent energy}

The energy spectrum $\C E(k,t)$ of isotropic turbulence can be described by the continuity equation %%
\BSE{cont}\be\label{contA}
\frac{\p \C E(k,t)}{\p t}+ \frac{\p \varepsilon(k,t)}{\p k}=0\,,\ee
where $\varepsilon(k,t)$ is the turbulent energy flux over scales. In the stationary case this equation simplifies to the requirement of the constancy of the energy flux in the  {so-called inertial interval,  where both energy pumping and energy dissipation can be neglected}:
\be\label{contB}\varepsilon(k)=\varepsilon\ .
\ee \ese%%
In order to describe a stationary spectrum $\C E(k)$ one needs to know how $\varepsilon(k)$ depends on  $\C E(k)$.  For  simplicity in this paper we will use reasonably simple differential models, that describe  the turbulent energy cascades of hydrodynamic (HD) and Kelvin wave (KW) turbulence at least   qualitatively and sometimes even semi-quantitative.

The first differential equation model for HD was first proposed by Leith in 1967 \cite{Leith67} and was recently studied in \cite{Nazar-Leith}:%%
\begin{equation}%%
  \label{Leith-67}%%
  \varepsilon\Sb{HD}(k) = -{1\over 8} \, \sqrt{k^{11} \C E\Sb{HD}(k)}\  {d\,  \over d k} \frac{\C E\Sb{HD}(k) }{  k^2} \ .
\end{equation}
Here $\varepsilon\Sb{HD}(k)$ is the energy flux carried by   HD turbulence. {For \Eq{K41-spectrum}, the factor $\frac18$ reproduces a numerical coefficient that reasonably fits the experimentally observed value of the Kolmogorov constant}.

Generic HD spectrum with a constant energy flux was found in \cite{Nazar-Leith} as a solution to the equation ~$\varepsilon\Sb{HD}(k) = \varepsilon = \mathrm{const}$: %%
\BSE{K41}\begin{equation}%%
  \label{K41-modif}%%
  \C E\Sb{HD}(k) = k^2 \left[ \frac{24\,\varepsilon }{11\, k^{11/2}} + \Big(\frac{T}{\pi \rho}\Big)^{3/2}\right]^{2/3}\ .%%
\end{equation}
The large $k$ range describes a thermalized  part of the spectrum
with equipartition of energy characterized by an effective
temperature $T$, namely, $T/2$ of energy per a degree of freedom,
thus, $ \C E_k = T k^2 \big/ \pi \rho$. At
low $k$, \Eq{K41-modif} coincides with the K41 spectrum:%%
\begin{equation}%%
  \label{K41-spectrum}%%
  \C E\Sb{HD}(k) = \left(24/11\right)^{2/3}\, \varepsilon^{2/3}\,  k^{-5/3}\,.
\end{equation}\ese

For Kelvin turbulence, the differential approximation model was suggested in \cite{Naz_kelvin}.
In a way similar to \Eq{Leith-67},  we suggest here a differential approximation for the energy flux,
carried by the Kelvin waves:
\begin{equation}
  \label{dKW}%%
  \varepsilon\Sb{KW}(k) = -\frac{5}{7}\frac{( k \ell)^8 \C E\Sb{KW}^4(k) }{ \Lambda ^5\kappa^7}\,\frac{d\C E\Sb{KW}(k) }{d k} \ .
\end{equation}
Note that this form is slightly less general than the one of \cite{Naz_kelvin} because it does not
take into account conservation of the waveaction. However, it is simpler which allows a more detailed
analytical treatment.

In the stationary case, equation ~$\varepsilon\Sb{KW}(k) = \varepsilon = \mathrm{const}$
has the solution
\BSE{KW}\begin{equation}
  \label{dKW-sol}%%
  \C E\Sb{KW}(k) = \Big[ \frac{ \Lambda^5\kappa^7}{\ell^8}\frac{\varepsilon}{k^{7}}  + \Big(\frac{T}{\pi \rho}\Big)^5 \Big]^{1/5}\ .
\end{equation}
This solution changes from KW-spectrum for small $k$:
 \begin{equation}%%
  \label{KW-spectrum}%%
  \C E\Sb{KW}(k) \simeq \Lambda \big(  \kappa^7 \varepsilon \big  / \ell^8 \big)^{1\!/5}\, k^{-7/5}\,,
\end{equation}
to the thermodynamically equilibrium solution with equipartition of energy  (Rayleigh-Jeans spectrum)
 \BE{KWa}\C E\Sb{KW}(k) = T /\pi \rho\,,
  \ee\ese
  for large $k$. The factor $-\frac{5}{7}$ in \Eq{dKW} is chosen such to reproduce in \Eq{KW-spectrum} the  numerical coefficient equal to unity. The  actual value of  this factor  is still not established with a reasonable accuracy, see, e.g. \cite{KS04}.

%Notice, that a more sophisticated model for $\varepsilon\Sb{KW}(k)$ with three differential operators $\frac{d}{dk}$   was proposed by Nazarenko in \cite{Nazar-KW}. Besides the conservation of energy this model accounts for the conservation of action (total particles number) and has two additional solutions: i) with constant action flux, and ii) with equipartition of action. We think that these solutions are not important in our problem and therefore for simplicity of further analysis we will use the more simple differential model~\eq{dKW}.

When the eddy (HD) and KW turbulence coexist, both models should work together in such a manner that for small $k$  the HD spectrum should be recovered, while  for   large $k$  the KW spectrum should be the only one:
\begin{eqnarray}
\C E(k) = \left\{
    \begin{array}{ll}
      \C E\Sb{HD}(k)\,, \qquad & k\ll 1/\ell \,, \\
      \C E\Sb{KW}(k)\,, \qquad & k\gg 1/\ell \ .
    \end{array}
  \right.
\end{eqnarray}
A way to reach this physical requirement is presented in the following section.

\subsection{A unified model for the total eddy-wave energy flux}
 A relatively simple  model of turbulence with two types of motions, random eddies and Kelvin waves,
is as follows.
% that we can imagine,  is to say,
The two types of motion coexist and interact in the extended  crossover  range in the following sense:
\paragraph{The total turbulent energy density} $\C E(k)$ and the total energy flux over scales, $\ve(k)$, consist of two respective parts:
\BSE{mod1}
\BEA{mod1-en} \C E(k)&=&\C E\Sb{HD}(k)+\C E\Sb{KW}(k)\,, \\
\ve (k)&=&\~\ve\Sb{HD} (k)+\~\ve\Sb{KW}  (k)\ ; \label{mod1-fl}
\eea\ese
where energy fluxes $\~\ve\Sb{HD} (k)= \ve\Sb{HD} (k)+ \ve\Sb{HD } \Sp{KW}$ and
$\~\ve\Sb{KW} (k)= \ve\Sb{KW} (k)+ \ve\Sb{KW} \Sp{HD}(k)$ have additional contributions $\ve\Sb{HD } \Sp{KW}(k) $ and $\ve\Sb{KW} \Sp{HD}(k)$ that originate from influence of KW on the HD-energy flux and vise versa.

 \paragraph{Continuity equations~\eq{contA} for the energy densities} have to be supplemented   by additional terms $\pm F(k)$, that describe energy exchange between two types of motion:
\BSE{mod2}
\BEA{mod2-HD} \frac{\p \C E\Sb{HD}(k,t)}{\p t}+ \frac{\p \~\varepsilon\Sb{HD}(k,t)}{\p k}&=&-F(k,t)\,,  \\
 \frac{\p \C E\Sb{KW}(k,t)}{\p t}+ \frac{\p \~\varepsilon\Sb{KW}(k,t)}{\p k}&=&F(k,t)\,, \label{mod2-KW}
\eea\ese

 \paragraph{Cross-contributions to the energy fluxes,} $\ve\Sb{HD } \Sp{KW}(k)$  and $\ve\Sb{KW } \Sp{HD}(k)$  are modeled in the  linear approximation with respect of the  influential energies (i.e. the HD energy influencing the KW flux and vice versa) :
 \BSE{fl}\BEA{flA}\ve\Sb{HD } \Sp{KW}(k) =\C D\Sb {HD} \{\C E\Sb {HD }\} \  d \, [\C E\Sb {KW}(k)/k_*^2] /d k^2 \,, \\ \label{flB}\ve\Sp{KW} \Sb{KW}(k)  =\C D\Sb {KW} \{\C E\Sb {KW}\} \ d \,  [\C E\Sb {HD}(k)/k^2] /d k^2\,,
 \eea\ese%%
 with some wave-vector $k_*$ which will be clarified later.
 Differential form of these contributions follows from physical hypothesis that these terms should disappear (or became much smaller and can be neglected) when the influential subsystem is in thermodynamical equilibrium, i.e. when $\C E\Sb {HD}\propto k^2$ and  $\C E\Sb {KW}\propto k^0=$const.    Functionals of the corresponding energies,  $\C D_{\dots}\{\dots\}$, will be modeled  by dimensional reasoning exactly in the way, how equations~\eq{Leith-67} and \eq{dKW} for the fluxes have been formulated. Resulting equations for $A$ can be written in the form:
 \BSE{m1}\BEA{m1A}
 \C D\Sb {HD} \{\C E\Sb {HD }\} &=& C\Sb{HD}\, \sqrt{k^{11} \C E\Sb{HD}(k)}\,, \\
 \label{m1B} \C D\Sb {KW} \{\C E\Sb {KW}\} &=& C\Sb{KW}(k\ell)\, k^{2}_*\C E\Sb{KW}^4(k)\kappa^{-7}\,,
 \eea\ese
where $C\Sb{HD}$ is a dimensionless parameter and   $C\Sb{KW}(k\ell)$ is a dimensionless function of $k\ell$, that will be chosen below in  \Eq{mod5}.

\paragraph{The energy distribution between the counterpart components} depends only on $k$ and for simplicity is assumed to be independent of the level of turbulence excitations:
\BSE{mod3}
\BEA{mod3-HD}   \C E\Sb{HD}(k,t) &=& g(k\ell )  \C E (k,t)\,,  \\
  \C E\Sb{KW}(k,t) &=& [1-g(k\ell)]  \C E (k,t)\,,\label{mod3-KW}
\eea\ese
were we introduced (only) $k\ell$-dependent blending function $g(k\ell)$ which will be explained below.

\paragraph{\label{p:1} Resulting model for the total energy flux \ $\ve(k)$.}
Adding the two \Eqs{mod2} and using \Eqs{mod1}, one yields the continuity \Eq{contA} in which $\ve(k)$ is given by \Eq{mod1-fl}. In such a way the unknown function $F(k,t)$ disappears from the game.   Together with \Eqs{Leith-67}, \eq{dKW}, \eq{mod1-fl}, \eq{fl}, \eq{m1} and \eq{mod3} this finally gives:
\BEA{mod4}
  \ve(k) = - \Big\{ \!\!\!&&\!\!\! \frac 18 \sqrt{k^{11} g(k\ell)\C E (k)} \\  \nonumber%%
  &&\!\!\!\! + \frac{5}{7}\frac{(k\ell)^8k_*^2  [1-g(k\ell)]^4 \C E(k)^4}{\Lambda^5\kappa^7} \,\Big\}\times \\  %%
  &&\ \ \frac{d}{d k}\Big\{ \C E (k)\Big[ \frac{g (k\ell )}{k^2}+ \frac{1-g (k\ell)}{k_*^2}\Big] \Big\}\ . \nonumber
\eea
In the derivation of this equation we took
\BE{mod5}
C\Sb{HD}=- 1/ 8\,, \quad C\Sb{KW}(k\ell)= - 5 (k\ell)^8/7 \Lambda ^5\ .
\ee
Only with this choice the resulting \Eq{mod4} for $\ve(k)$ is proportional to $d [\C E(k)/k^2]/ dk$ and $\ve(k)$ vanishes in the thermodynamical equilibrium with $\C E(k)\propto k^2$, as one should expect.

 Equation~\eq{mod5} contains yet unknown blending function $g(k\ell)$ which will be discussed in the next section.

\subsection{\label{ss:sep} Separation  of the eddy and wave motions}

In order to find a qualitative form of the blending function we consider a system of locally (in the vicinity of some point $\B r_0$) near-parallel vortex lines, separated by mean distance $\ell$ and supply them by index $j$. Notice that in principle the same vortex line can go far away and
come close to $\B r_0$ several times. To avoid this problem one should assign the same vortex
line a different index $j$ if it leaves (or enters) the ball of radius $  \ell\sqrt \L$
centered at $\B r_0$. Each vortex line (with zero radius $a$) produces a velocity field $\B v_j(\B r)$, which can be found by the Biot-Savart Law~\eq{BS}.

 The total kinetic energy   $E =\frac12 \sum_{i,j} \<\B v_i\cdot \B v_j\> $ can be divided into two parts, $E=E_1+E_2$, where
\BE{eq1}  E_1\= \frac 12 \sum_j \< v_j^2\>\,, \quad  E_2\= \frac 12 \sum_{i\ne j}\< \B v_i\cdot \B v_j\> = \sum_{i< j}\< \B v_i\cdot \B v_j\>\ .
\ee
The same subdivision can be made also for the energy density in the (one-dimensional) $k$-space,  $\C E(k) = \C E_1(k) +\C E_2(k)$, with two terms, that can be found via $\B k$-Fourier components of the velocity fields $\B v_j(\B k)$ in the way, similar to \Eq{eq1}.
Now our idea is as follows: energy $\C E_1(k)$ is defined  by the form of the individual vortex lines, that is determined by the Kelvin waves, while energy  $\C E_2(k)$ depends on correlations in the form of different vortices, that produce collective, hydrodynamic type of motions.  Therefore $\C E_1(k)$ can be associated with the Kelvin wave energy, $\C E_1(k)\Rightarrow \C E\Sb{KW}(k)$, while $\C E_1(k)$ has to be associated with the
hydrodynamic energy,  $\C E_2(k)\Rightarrow \C E\Sb{HD}(k)$. This allows one to conclude that
\BE{eq2}
g(k\ell)= \big[1 + \C E_1(k)/\C E_2(k)\big]^{-1}\ .
\ee
The rest is technicalities    presented in Appendix~\ref{a:bl}, where we concluded that in practical calculations it is reasonable to   use analytical form $g(k\ell)$ of the blending function%%
\BSE{est5}
\begin{equation}
  \label{est5C}
  g(k\ell)  =  g_0\big[0.32\,\ln (\L+7.5)\ k\ell\big]\,,
\end{equation}
where
\begin{equation}
  \label{est5B}
  g_0(k\ell) = \Big[ 1+ \frac{(k\ell)^2\exp(  k\ell )}{4\pi (1+  k\ell)}\,\Big]^{-1}\ .
\end{equation}
\ese

%%\BSE{est5}%%
%%  \BEA{est5B} %%
%%    g_0(k\ell)&=& \left[ 1+ \frac{(k\ell)^2\exp(  k\ell )}{4\pi (1+  k\ell)}\right]^{-1}\,,\\ %%
%%    \label{est5C} %%
%%    g(k\ell)  &=&  g_0\big[0.32\,\ln (\L+7.5)\ k\ell\big]\ .
%%  \eea %%
%%\ese

\subsubsection{\label{sss:X-s}Comparison of various  crossover  scales}
With the proposed blending function we can introduce another   cross-over scale $k\sb{en}\ell $ by comparing  HD and KW energies,  at  $k\sb{en}\ell $ they are equal: $g(k\sb{en}\ell)=1/2$. As follows from \Eqs{est5}, $k\sb{en}\ell $ has very weak, logarithmical dependence on $\L$:
\BE{X2}
  k\sb{en}\ell \simeq   6.64 \big /  \ln (\L+7.5)\,,
\ee
 presented in the second column in Table.~\ref{t:1}.

First column of this table displays velocity-crossover scales, given by  \Eq{XB}: $k\sb{vel}\ell \simeq \sqrt{2/\Lambda}$, while in the column 3 one finds the amplitude crossover $k\sb{amp} \ell$, \Eq{Xamp}, scale at which the formally computed with the KW spectrum amplitude of Kelvin waves reaches intervortex distance.   The fourth column of Table.~\ref{t:1} displays flux-crossover scale, $k\sb{fl}\ell$ at which the contribution of the eddy- and Kelvin-wave turbulence to the energy flux in the $k$-space become equal. This scale is introduced below in Sec.~\ref{ss:dif}.

\begin{table}
\begin{tabular}{||c||c|c|c|c|c||}
\hline\hline
-- & 1&2&3&4&5\\ \hline
$\Lambda$  & \begin{tabular}{c}
                  $k\sb{vel }\ell$ \\
                  \Eq{XB} \end{tabular}
                  &\begin{tabular}{c}
                  $k\sb{en}\ell$ \\
                  \Eq{X2} \end{tabular}
                  & \begin{tabular}{c}
                  $k\sb{amp}\,\ell$ \\
                  \Eq{Xamp} \end{tabular}
                   & \begin{tabular}{c}
                  $k\sb{fl}\ell$ \\
                  Numerics  \end{tabular}
                   & \begin{tabular}{c}
                  $\epsilon\!\times\!\!10^{-3}$ \\
                  self-cons.  \end{tabular}
                   \\ \hline\hline
10      & 0.45    & 2.3   & 0.76  & 24.4 & 5.2      \\ \hline
30      & 0.26    & 1.8  & 0.64  & 23.6  & 2.1      \\ \hline
$10^2$  & 0.14    & 1.4  & 0.52  & 22.8 & 0.30      \\ \hline
$10^3$  & 0.045   & 0.96   & 0.35  & 22.5 & $1.2\!\times\!\!10^{-3}$      \\ \hline
\hline
\end{tabular}
\caption{\label{t:1}  Comparison of the energy-, velocity-, amplitude- and flux-crossover scales for typical experimental  values $\Lambda=10$ and 30  and unrealistically large values $\L=100$ and $10^3$. Values of $k\sb{fl}$ depend on $\epsilon$ defined in a self-consistent way, explained in \Sec{ss:sc} }
\end{table}

Important message is that  although  theoretically, in the limit $\Lambda\to \infty$ (see Appendix), $k\sb{en}\ll k\sb{amp}$, in the  region $10\lesssim \Lambda \lesssim  30$   these scales are close, moreover, due to numerical prefactors
$k\sb{en}> k\sb{amp}$. Therefore the fractal structure of the vortex lines, described in \Sec{ss:X}, does not appear which allows   us to  use  in the actual region  $10\lesssim \Lambda \lesssim  30$ the proposed simple blending function (\ref{est5B}).

\begin{figure*}
  \includegraphics[width=8 cm]{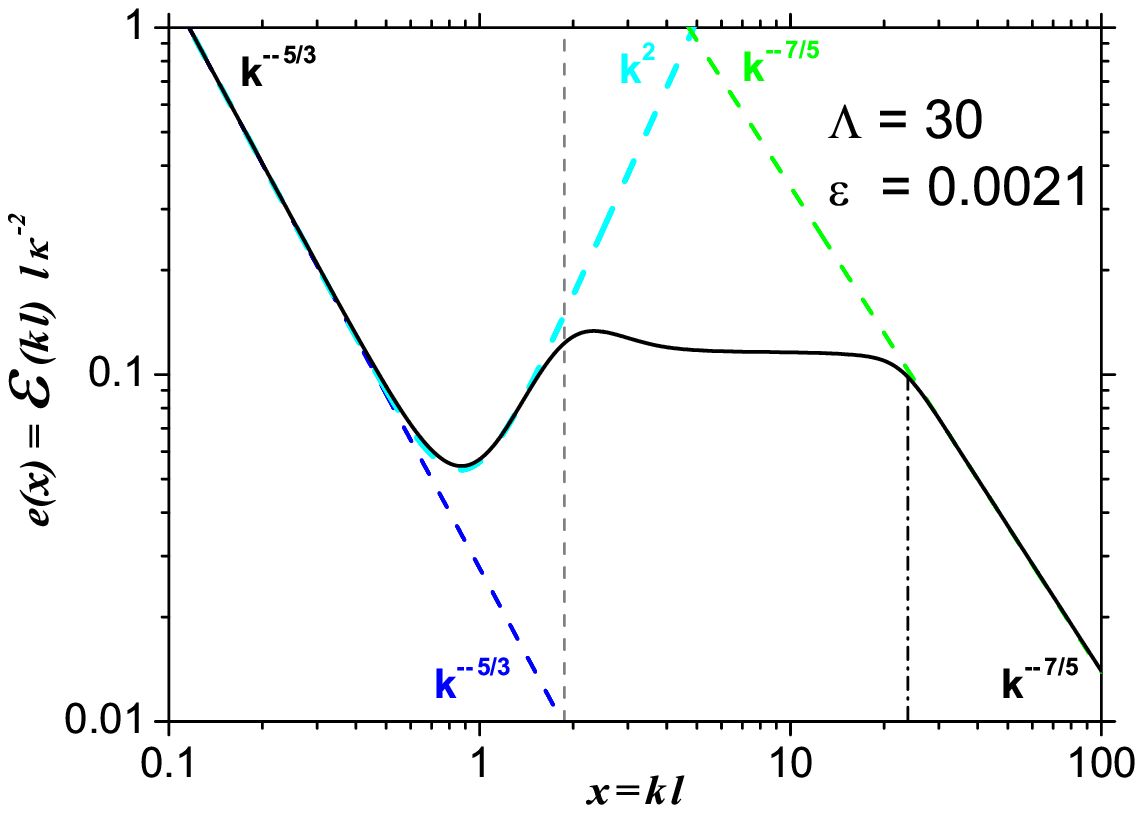}~
   \includegraphics[width=7.85 cm]{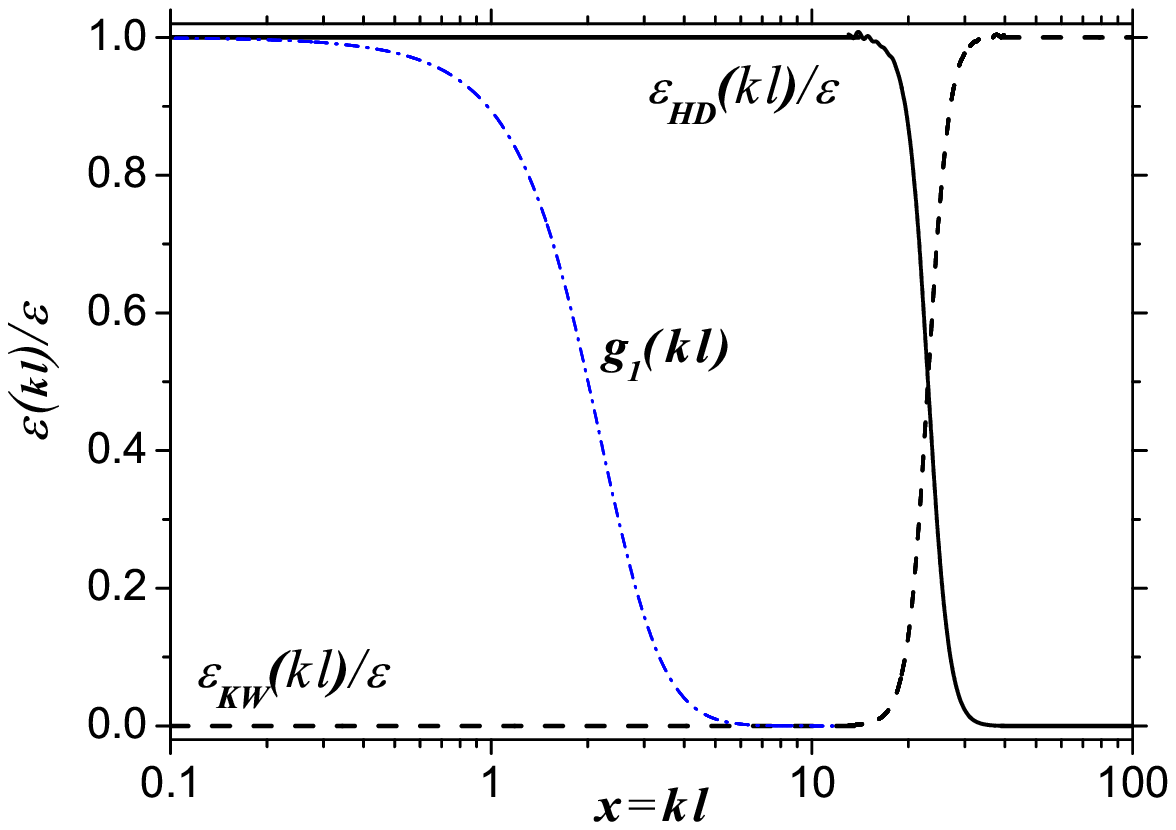}
  \caption{\label{f:e-typical}  \textbf{Left panel:} Solid (black) line represents typical total dimensionless energy spectrum $e(x)$ obtained by numerical solution of the ODE (\ref{ODE-ode}) with $\Lambda = 30$ and self-consistent value of $\epsilon = 0.0021$, found in \Sec{ss:sc}. Dashed (blue) line corresponds to the K41 energy spectrum $e\Sb{HD}(x)\propto x^{-5/3}$ with constant energy flux, dashed (cyan) line is general HD spectrum  $e\Sb{HD}(x, T\!  = \! 0.042)$,  dashed (green) line is the energy spectrum of Kelvin waves  $e\Sb{KW}(x)\propto x^{-7/5}$.   Vertical dashed (gray) line shows $x\Sb{en}$. Vertical dot-dashed (brown) line shows position $x\Sb{fl}$, where $\epsilon\Sb{HD}(x\Sb{fl}) = \epsilon\Sb{KW}(x\Sb{fl})$. \textbf{Right panel:} Partial energy fluxes $\epsilon\Sb{HD}(x)/\epsilon$ (solid line) and $\epsilon\Sb{KW}(x)/\epsilon$ (dashed line) obtained by numerical   solution of the ODE (\ref{ODE-ode}) with $\Lambda = 30$ and $\epsilon = 0.0021$. Dot-dashed (blue) line represents $g(x)$.}
\end{figure*}

\begin{figure*}
\includegraphics[width=8 cm]{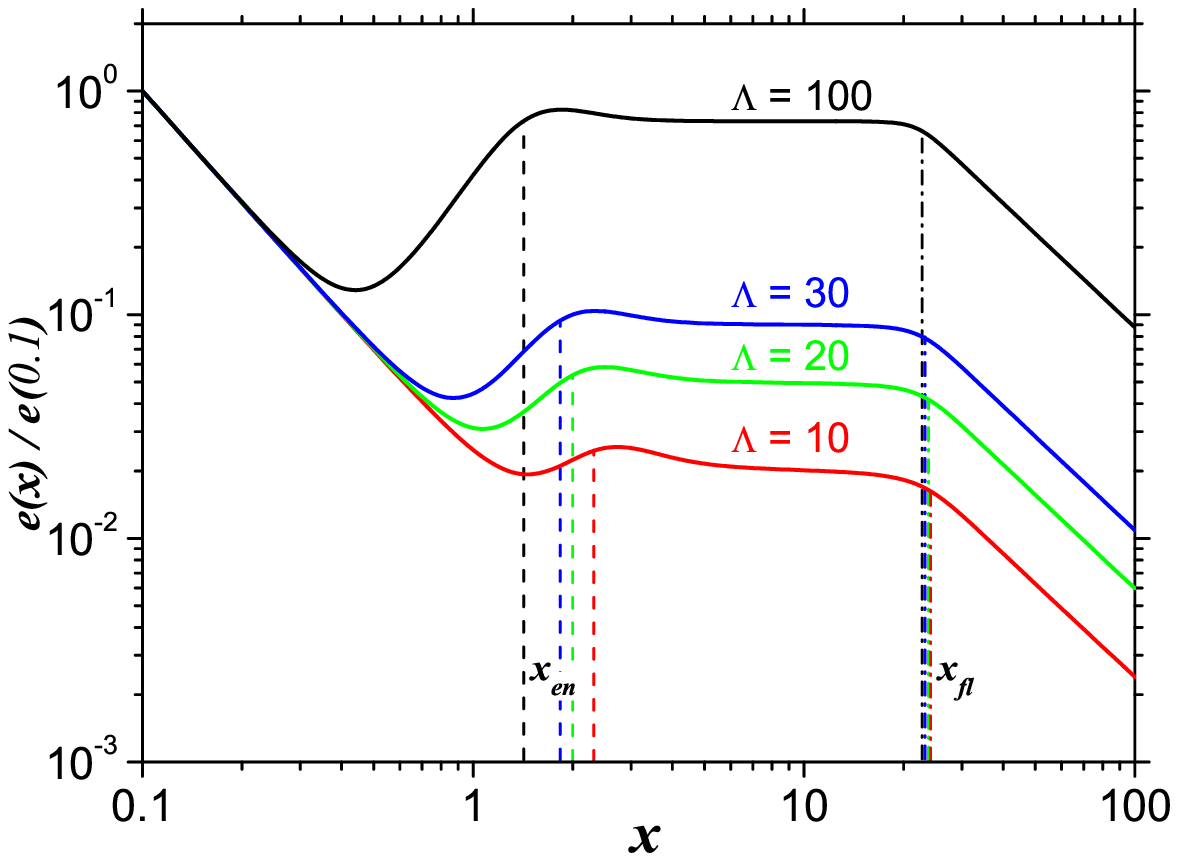}~
  \includegraphics[width=8 cm]{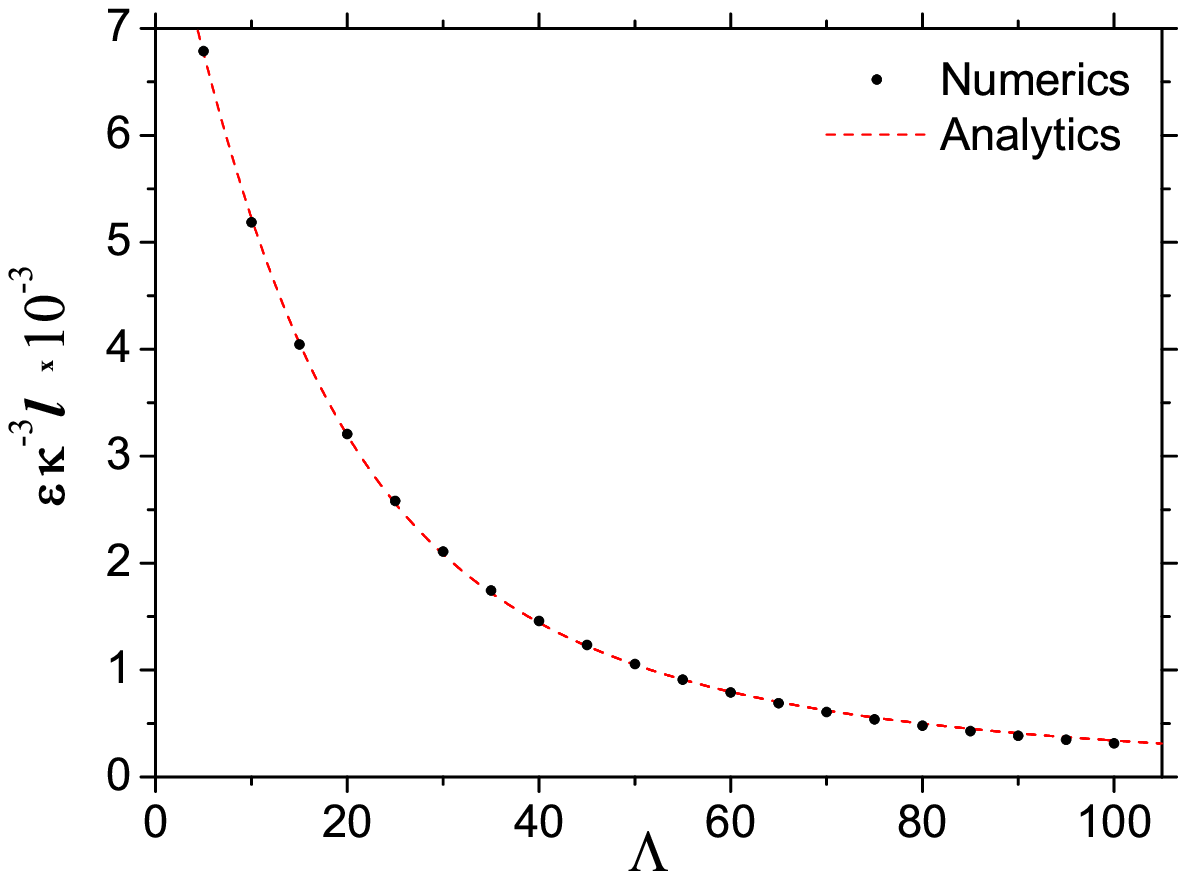}

  \caption{\label{f:exL30} Normalized non-dimensional energies $e(x)$ at $\Lambda =10, 20,  30$ and 100 (left) and $\L$-dependence of the self-consistent energy flux $\e$.    Vertical dashed (gray) line shows $x\sb{en}$.  Vertical dash-dotted lines indicate positions of $x\sb{fl}$ (in red, green and blue for corresponding $\epsilon$).  }
\end{figure*}

~
\subsection{\label{ss:dif} Turbulent energy spectra in the differential approximation}

\subsubsection{\label{sss:norm} Dimensionless representation}

At the end of the day we are left with solving the ODE (\ref{mod4}), where the blending function is $g(k\ell)$, \Eqs{est5B}.  At this stage of research it is reasonable to nondimensionalize physical quantities introducing%%
\begin{equation}
  \label{non-dim-vars}%%
  x = k\ell\,, \quad e(x) = \frac{\ell}{\kappa^{2}}\, {\C E(x)}\,, \quad \e = \frac{\ell^4}{\kappa^{3}}\,\ve\ .
\end{equation}
In particular, with this normalization the one-dimensional energy spectra  HD and KW,~\Eqs{K41} and \eq{KW-spectrum},    take the form
\BSE{sp}\begin{eqnarray}
  \label{f-HD}%%
 e\Sb{HD}(x)   &=& (24/11)^{2/3} \epsilon^{2/3} x^{-5/3}\,, \\ %%
  \label{f-HD-T}%%
  e\Sb{HD}(x,T) &=& x^2 \left[\frac{24}{11} \frac{\epsilon}{x^{11/2}} + T^{3/2}\right]^{2/3}\,, \\
   \label{f-KW}%%
  e\Sb{KW}(x) &=& \Lambda \epsilon^{1/5} x^{-7/5}\,,   %%%%
\end{eqnarray}\ese
where $T$ is non-dimensional temperature.
 And the ODE~\eq{mod4} to solve [with the boundary condition $e(x)\to e\Sb{KW}(x)$ for $x\gg 1$] becomes %%
\begin{eqnarray}
  \label{ODE-ode}%%
  \epsilon = -\Big\{ \!\!\!&&\!\!\!\! \frac{1}{8}\sqrt{x^{11} g(x) e(x)}  \\ \nonumber %%
   &&\!\!\! + \frac{5}{7}\, \frac{x^{8}x\sb{en}^2}{\Lambda^{5}}  \big[ 1 - g(x) \big]^4 e^4(x) \Big\}\times \\ \nonumber %%
   && \ \ \ \frac{d}{d x}\,e(x)\Big[\frac{g(x)}{x^2} + \frac{1-g(x)}{x\sb{en}^2}\Big]\ .
\end{eqnarray}
Here we made  the natural choice that the crossover scale $k_*$ between two types of thermodynamic equilibrium is  $k\sb{en}$, the scale where energies of two types of motion are the same.

\subsubsection{\label{sss:tipe} $\L$-dependence of the energy spectra}

 An instructive solution $e(x)$ with $\Lambda = 30$ and   $\epsilon = 2.1\, 10^{-3}$ is shown
 in \Fig{f:e-typical}, left, as a (black) solid line. One sees that this solution for $x\lesssim x\sb{en}\simeq 1.83 $   follows the thermolized HD spectrum $e\Sb{HD}(x,T)$  [given by \Eq{f-HD-T} with properly chosen $T$] shown as a dotted (cyan) line.
 An important observation is  that the pseudo-thermolized part of the spectrum is very pronounced in the region $x \gtrsim 0.5 $ where it is very different  from   the K41 spectra of HD turbulence $e\Sb{HD}(x)\propto x^{-5/3}$ shown  as a dashed (blue) line.  For $x  \gtrsim x\sb {fl}$ the solution practically coincides with the  pure KW spectrum $ e\Sb{KW}(x)$, \eq{f-KW} shown  as a dashed (green) line. Important, that the crossover scale $x\sb {fl}\simeq 23.6  $, at which the total energy flux consists of 50\% of HD- and 50\% of KW-fluxes, is much larger that $x\sb{en}\simeq 1.83$, at which a half of the total energy is carried by HD and half by KW motions.  To make this evident we plotted in \Fig{f:e-typical}, right, the partial HD- and KW-energy fluxes vs. $x$. They become equal at $x\sb{fl}$, which  for $\Lambda = 100$ and $\epsilon = 3\, 10^{-4}$ is around 23.6.

 In the intermediate region $x\sb{en}<x<x\sb{fl}$ the energy consists mostly of the KW energy, while the energy flux is carried mostly by the HD motions.  Explanation to this observation is simple: as follows from \Eq{ODE-ode} the HD motions are more effective (in factor $\sim \L^5 / x^{9/2}$)  in support of the energy flux then the KW turbulence. Because the main part of the energy flux is taken by the HD motions, the KW energy spectrum (and therefore the total one) is close to the flux-less KW-solution: thermodynamic equilibrium~\eq{KWa}, $\C E\Sb{KW}=$const. For $x> x\sb{fl}$, both the energy and the energy flux are carried by the KW motions. Therefore the total energy spectrum coincides with the KW cascade solution.
 For larger values of $\e$ the flux-crossover scale goes to the smaller values of $k$, see \Fig{f:exL30}, remaining nevertheless larger than $k\sb{en}$.

 For smaller value of $\L$ qualitative behavior of $\C E(x)$ remains the same, just different parts of the spectra (with larger values of self-consistent values of $\e$) become less pronounced.

\subsubsection{\label{ss:sc} Self-consistent estimate  of the dimensionless energy flux}
Energy spectra, \Figs{f:e-typical} and \ref{f:exL30},  which we obtained in the proposed differential approximation  are   quite similar to those suggested under the assumption of sharp crossover, see  Fig.~1 in  \Ref{lnr}: for small $k$ they coincide with the HD spectrum, including the bottleneck part with (almost) thermalized part $\C E\propto k^2$ , while for large $k$ the spectrum follows the KW spectrum $\propto k^{-7/5}$.
The only difference is that with sharp crossover the thermalized part of the HD-spectrum is matched  with KW-spectrum at some $k=k_*$, while  in the differential approximation with a smooth blending function there is essential intermediate region (about one decade)  $k\sb {en}<k<k\sb{fl}$, with  (almost) KW-thermalized part $\C E=$const.  This leads to an essential difference in the estimates of the vorticity $\< \o^2\>$ and as a result in the estimates of the effective viscosity $\nu^{\prime}$, a parameter that can be measured (implicitly) in experiments.  Indeed, in the models with the sharp crossover one estimates $\<\o^2\>$ in \Eq{est-our} as $k_*^3 \C E\Sb{HD}(k_*)$ and then
equates $\C E\Sb{HD}\simeq \C E\Sb{KW}$, because in this model the HD-energy ``transforms" into the KW-energy at the position of the sharp crossover.

 In the ``continuous'' model presented above one has to account for  a wide region between $k\sb{en}$ and $k\sb{fl}$, where the flux is supported by the HD turbulence, while the energy is dominated by the Kelvin waves. Therefore  $\< \o^2\>$ can be estimated similarly to \Eq{est-our} as $2\int k^2 \C E\Sb{HD}(k) dk \simeq k\sb{en}^3 \C E\Sb{HD}(k\sb{en})$, but now
$\C E\Sb{HD}(k\sb{en})$ cannot be estimated based on the $-7/5$ Kozik-Svistunov spectrum because,
%as $\C E\Sb{KW}(k\sb{en})$ because $\C E\Sb{HD}(k\sb{en}) \ll \C E\Sb{KW}(k\sb{en})$,
as one sees in \Fig{f:exL30}, $\C E\Sb{HD}$ is much lower at this point. As a result, at the same energy flux the  rms. vorticity $\sqrt{\<|\o|^2\>}$ occurs to be essentially smaller then that in the sharp-crossover models and the effective viscosity is larger. In our approach the rms vorticity can be found more accurately by numerical calculation of the integral in \Eq{est-our} with  spectrum $\C E\Sb{HD}(k)\simeq g(k\ell)\C E (k)$, where the blending function is given by \Eqs{est5}:
\BSE{est6}\BE{est6A}
\<|\o|^2\>= 2 \int _{k\sb{min}}^\infty k^2 g(k\ell)\C E (k)\, dk\,,
\ee
where $k\sb{min}$ is the lower cutoff of the inertial interval.
Using relation $\<|\o|^2\>= \kappa^2/ \ell^4$ and normalization~\eq{non-dim-vars} one finds from \Eq{est6A}  in the limit $k\sb{min}\to 0$:
\BE{est6B}
1= 2  \int_0^\infty x^2 g(x) e (x)\, dx\ .
\ee
\ese
Due to $\e$ dependence of the energy spectrum $e$, this relation gives self-consistent a estimate of the dimensionless energy flux $\e$, which is according to \Eqs{nu-p} and \eq{non-dim-vars} is nothing else, but $\nu^\prime/\kappa$. Resulting dependence $\e$ vs. $\L$ is shown in \Fig{f:exL30}, right, as solid line. For convenience we approximate this dependence (in the actual interval $\L<100$) analytically:
\BE{e-an}
\e=\frac{\nu^\prime}{\kappa}= \frac{8.65}{10^3 + 45.8 \L + 1.98 \L^2}\,,
\ee
shown in  \Fig{f:exL30}, right, as dashed line.  Equation~\eq{e-an} reproduces numerical dependence $\e(\L)$ with accuracy better than 1.5\% for $\Lambda < 50$ and better than 8\%
 for $50< \Lambda < 100$.

 Notice, that found values of $\e=\nu^\prime/\kappa$ for $\L=30$ is 0.021 which is  quite  close to the experimentally reported value  $\nu^\prime\simeq 0.003 \kappa$ in $^4$He experiments at low temperatures.
A relationship between out model and experiments will be discussed below.

\subsection{\label{ss:decay}Decay of quantum turbulence with the
bottleneck energy accumulation}

Having in mind experiments with decaying superfluid turbulence, like the ones
in~\cite{Manchester-exp},
it is important to discuss how the bottleneck energy accumulation
influences the decay of energy and vorticity in time. For this, we
divide the total HD energy  %%
\BSE{en}\BE{enA}E\Sb{HD}=\int _{k\sb{min}}^\infty  dk\, \C E \Sb{HD}(k)%
  \ee%%
  into a sum of two parts:
  \BE{enA} E\Sb{HD}= E \Sb{HD}\Sp{K41}+ E \Sb{HD}\Sp{TE}\,,
  \ee
  the energy $E \Sb{HD}\Sp{K41}$ associated with the K41 part of
energy spectra $\C E\Sb{HD}\propto k^{-5/3}$,  and the energy
$E\Sp{TE}\Sb{HD}$ associated with the thermodynamic equilibrium (TE)
part of the spectrum  $\C E\Sb{HD}\propto k^{2}$. For our model:
\BEA{enB}%%%
   E\Sb{HD}\Sp{K41}&=&\int ^{k\Sb{TE}}_{k\sb{min}} dk\,  \C E\Sb{HD}
(k)\, \,,\\ \label{enD}
E \Sb{HD} \Sp{TE}  &=& \int ^\infty_{k\Sb{TE}}  dk\, \C E \Sb{HD}(k)\,,
  \eea\ese
  where  $k\Sb{TE}$ is the crossover scale between K41 and TE parts of
the energy spectra corresponding to the position where $\C E \Sb{HD}(k)$
is minimal. For $\L=30$, $k\Sb{TE}\approx 1/\ell$, see
\Fig{f:e-typical}, left. The K41-energy, $E \Sb{HD}\Sp{K41}$, is
dominated by the outer region of the $k$-space, $k \gtrsim k\sb{min}$,
while the TE-energy is determined by effectively the largest $k\simeq
k\sb{en}$ of the HD motions (\Fig{f:e-typical}, left). In the
experiment~\cite{Manchester-exp}, $k\sb{min}$ is below 1 cm$^{-1}$,
which is less than $k\sb{en} \simeq 2/\ell$ by one or two orders of
magnitude. Then, the experiment~\cite{Manchester-exp} allows to estimate
the ratio $E\Sp{TE}\Sb{HD}/E\Sp{K41}\Sb{HD}$ in the proposed framework,
and it varies from a few percents for small times to about 15-20 \% at
the latest  times of the decay measurements.  For us this means that with
an acceptable accuracy one can neglect the contribution of
$E\Sp{TE}\Sb{HD}$ in Eq.~(\ref{enA}).

   Moreover, even when kept in  Eq.~(\ref{enA}), the energy
$E\Sp{TE}\Sb{HD}$ does not appreciably affect on the decay rate of
$E\Sp{K41}\Sb{HD}$ energy due to a large scale separation ($k\sb{min}
\ll k\Sb{en}$). The decay rate of $E\Sp{K41}\Sb{HD}$  is determined by
the energy flux $\ve = -dE\Sp{K41}\Sb{HD}/dt$ at the scale of the energy
pumping, i.e. at the outer scale $k=k\sb{min}$. The flux itself is
proportional to $\C E\Sb{HD}(k)^{3/2}\sim  (E\Sb{HD}\Sp{K41}/k\sb{min}
)^{3/2}$.  For systems with the time independent   $k\sb{min}$, as it is
in~\cite{Manchester-exp}, this gives the well known result for the
late-time free-decaying HD turbulence:
\BSE{laws}\BE{lawsA}
E\Sb{HD}\Sp{K41}(t)\propto t^{-2}\,,%%
\ee%%
and the time-evolution of the energy flux
  \BE{lawsB}
  \ve(t)\propto  t^{-3}\ .%%
  \ee%%
According to \Eq{non-dim-vars}, ~$\ve(t)=\e \kappa^3/\ell^{4}(t)$ with
the time-independent self-consistent dimensionless energy flux $\e$,
which depends only on $\L$.  This gives $\ell(t) \propto t^{3/4}$.
Therefore, the vortex line density  must decay in the standard manner:
%%\cite{Manchester-exp}:
  \BE{stan}
  L=1/\ell^2  \propto t^{-3/2}\,,
  \ee
  in spite of the   accumulation of energy $E\Sb{HD}\Sp{TE}$ near the
crossover scale $k\Sb{en}$.

  Notice, that  energy $E\Sb{HD}\Sp{TE}(t)$ decays slower than $
E\Sb{HD}\Sp{K41}(t)\propto t^{-2}$.
  Indeed, in our model the dimensionless  energy $E\Sb{HD}\Sp{TE} \ell^2/\kappa^2$ (cf. \Eq{non-dim-vars}) is dominated by the time independent scale $x\sb{en}$ and, hence, by itself is time independent. Therefore,
  \BE{lawsD}
   E\Sb{HD}\Sp{TE}(t)\propto \ell^{-2} \propto t^{-3/2}\ .
  \ee
  \ese

  One concludes  that  $E\Sb{HD}\Sp{TE}$  energy is ``decoupled" from
the decay process of $E\Sb{HD}\Sp{K41}$
  energy and does not affect  the   decay law~\eq{stan} of the vortex
line density
  until to the very late stage of the decay, when the intervortex
distance approaches the outer
  scale of turbulence and the entire model fails.

\section*{ \label{s:sum}  Summary and Discussion}

In this paper, we revised the theory of the bottleneck crossover from the classical K41 cascade to the Kelvin wave cascade.
In its previous form, transition from the eddy to the wave cascades was assumed to occur sharply at the scale $\ell$.
The simple fact that the wave interactions are less efficient for the turbulent cascade than the hydrodynamic eddies
immediately yields prediction for the bottleneck accumulation near the crossover scale.
However, the bottleneck strength is rather sensitive to the details of the crossover region.
In the present paper, we take into account that there exists a finite range where eddies and waves coexist and affect
each other, making the crossover more gradual. As a result, the bottleneck in such a case is milder than
in the model with the sharp crossover. To model the gradual transition range, we have employed a simplified
turbulence model which is based on the differential approximation models of Leith type for the
HD and KW components. Importantly, this model allows to make predictions for the realistic experimental
values $\L$ in the range from 10 to 30, rather than making asymptotical predictions for the
case $\L \to \infty$. This appears to be important because, e.g., the asymptotic theory gives
$k\sb{amp} \gg k\sb{en}$ whereas for $\L$ in the range from 10 to 30 we have
$k\sb{amp} \lesssim k\sb{en}$. As a result, for the experimentally important situations there is no range
with equipartition of the eddy and the wave energies given by (\ref{bal}).
For similar reasons, the theory of crossover \cite{KS07} which fits three asymptotic ranges into
a single decade of scales is rather unrealistic (leaving aside the issue about the role of reconnections which
we mentioned before).

One may experience some problems trying to imagine any HD components at $k\ell > 2\pi$, where the wavelength becomes larger than the intervortex distance, and even come to an idea that $g(k\ell)$ must become zero sharply at $k\ell = 2\pi$, or generally, $k\ell \sim 1$.
%Similarly, not so long ago it was difficult to imagine, how an electron can go simultaneously through two holes.
Our model is based on a (reasonable) hypothesis that the  HD motions are identified with the coherent part of different vortex line motions. In such an approach there is no formal limitation for the value of $k$ from above. At $k\ell>1$ the velocity produced by $k$-distortion  of a given vortex line in the position of another $\ell$-separated line, which is the reason for correlations in their motions, decays exponentially with $k\ell$. That is why our blending function, which measures the fraction of the HD motions, decays exponentially with $k\ell$. The actual hypothesis in this place is  that even when $g(k\ell)\ll 1$ the nonlinear energy flux carried by the HD motions is governed by the same equations as for the pure HD motions when $g(k\ell)=1$.  We believe this is a step forward in comparison with the oversimplified scenario of  sharp crossover which, by the way, is a  limiting case for our model where the blending function is just a unit-step function $g(k) = \Theta(k\sb{dim}-k)$.

The found value of $\e=\nu^\prime/\kappa$ for $\L=30$ is 0.021 which is  quite  close to the experimentally reported value  $\nu^\prime\simeq 0.003 \kappa$ in $^4$He experiments at low temperatures. Having in mind that our model does not contain fitting parameters, optimists  can consider this agreement as more than satisfactory. On the other hand, pessimists  can consider any agreement with just one number as accidental. Realists should recall that the suggested   model is based on a hypothesis of blending function, which was estimated without taking into the account the vectorial structure of the velocity field and, moreover, includes very important (step-like) assumption about the pair-distribution function of the vortex positions which allows one to estimate sum~\eq{est4A} as integral~\eq{est4B}. Our feeling is that these approximations do not affect the results too much and a good agreement between the model and experimental values of $\nu^\prime$  supports the suggested model.   Notice that our model predicts not only the value of $\nu^\prime$ but the entire energy spectrum, which consists of four parts: K41 HD energy spectrum with constant energy flux, $\C E\propto k^{-5/3}$, a HD equilibrium $\C E\propto k^{2}$, a KW equilibrium $\C E\simeq$ const and a KW-spectrum with constant energy flux, $\C E\propto k^{-7/5}$.  This very definite qualitative prediction calls for more detailed experimental and numerical study of the superfluid turbulence, which, as we believe, will support our model.

\section*{Acknowledgements}
This work has been partially supported   by  the Transnational Access Programme at RISC-Linz, funded by the European Commission Framework 6 Programme for Integrated Infrastructures Initiatives under the project SCIEnce (Contract No. 026133).

\appendix

\section{\label{a:bl}
Estimation of the HD-KW blending function}
\subsection{ Estimation of the velocity field, induced by the vortex distortion }

 To estimate the HD-KW blending function $g(k\ell)$, given by \Eq{eq2} we consider a vortex line slightly distorted, say in  $y$-direction, by a sinus with a small amplitude $A$, running along $z$-axis with the $k$-vector $k$. Then $d\bm \ell = (0,\, Ak \cos(k z+\phi),\, 1) dz $, where $\phi$ is an arbitrary phase. Each line produces velocity that can be found via Bio-Savart Law:
 \begin{eqnarray}%%
  \label{BS}%%
  \bm v_j(\bm r) = \frac{\kappa}{4\pi}\int_{-\infty}^{+\infty}{\frac{d\bm \ell_j \times \bm s_j}{s_j^3}}\,,
\end{eqnarray}
where $\B s_j=\B r-\B r_j$ with $\B r_j$ being the radius-vector pointing to the  $\B \ell_j$ -- the length element along the $j$-th vortex line and.  The total mean density of the kinetic energy per unit mass is $E=\frac12 \< |V|^2\>$, where $\B V(\B r)=\sum_j \B v_j(\B r)$ is the total velocity field, and one understands $\<\dots\>$ as the averaging with respect of the (random) vortex-line positions.

Using \Eq{BS} we can find the resulting   velocity   (for simplicity) in the $ZY$-plane at distance $R$ from the line. This velocity has only  one component, $v_x$, and it is given by:%%
\begin{eqnarray}%%
  \label{sin-vor}%%
  v_x \!=\! \frac{\kappa}{4\pi}\!\!\int_{-\infty}^{\infty}\!\!{\frac{R\! -\! A \sin{\!(k z\! +\! \phi)} + Ak z \cos(k z\! +\! \phi)}{\left\{ z^2 + \left[ R - A \sin{(k z + \phi)} \right]^2 \right\}^{3/2}}\, dz}\,.~
\end{eqnarray}
Computing the contribution to the amplitude of the velocity variations proportional to the distortion amplitude $A$, one finds magnitude [i.e. factor in front of $\cos(kz+\phi)$] of the velocity fluctuations:
\BSE{appr}\begin{eqnarray}%%
  \label{dv-Bess}%%
  \delta v (R) &=&  \frac{\kappa\,A}{2\pi R^2}\, kR \times \\ \nonumber
   \!\!\!\!&&\  \Big \{ kR[ K_2(kR) \!-\! K_0(kR)] \!-\! K_1(kR)\Big \},\ \ \Rightarrow \\ \label{apprB}
   \delta v(R)  &\simeq&  \frac{\kappa\,A}{2\pi R^2}\, \sqrt{1 +\frac{\pi}{2}\, kR\,}\, \exp (-kR) \,,
\end{eqnarray}
where  $K_n $ are modified Bessel functions of the second kind.
Interesting, that a simple interpolation~\eq{apprB}  reproduces the exact result~\eq{dv-Bess} with an accuracy better than $2\%$.

Equations~\eq{appr} assume that the distance $R$ to the vortex line is large enough to be able to neglect finiteness of the core radius $a$. In the opposite limit, $R\ll a$, one can use an equation for the self-induced velocity, in which the core radius is hidden in the parameter $\Lambda=\ln (\ell/a)$:
\BE{si}
\d v\sb {si}= A \Lambda k^2\k / 4\pi  \ .
 \ee%%
 Considering this equation formally as a limit for the velocity when $R\to 0$, we suggest an interpolation formula approximately valid for any $R$:
\begin{eqnarray}%%
  \label{dv-int2}%%
  \delta v_k(R) \!\!&\simeq&\!\! \frac{\kappa A}{2\pi}\, \frac{\Lambda k^2\, e^{-kR}}{2 +\Lambda (k R)^2}\, \sqrt{1 +\frac{\pi}{2}\, kR\,}\, ,~~~~~~
\end{eqnarray}\ese
where subscript  ``~$_k$~"  reminds that this velocity is induced by vortex line, distorted by $\sin(k r)$.

\subsection{\label{sss:prel}Velocity crossover scale}
 A way to estimate the eddy-wave crossover scale is to compare the self-induced velocity filed with the velocity field produced by a different neighboring vortex line. For this goal we consider simple geometry with  two sin-disturbed parallel vortex lines separated by distance $\ell$  They ``start to feel" each other when $\delta v \simeq \delta v\Sp{SI}$, i.e.:%%
\BE{X1A}%%
  \delta v \simeq \delta v\Sp{SI} \ \ \Leftrightarrow \ \ %%
  \, \sqrt{4 +2{\pi}\,k\sb{vel}\ell\,}\, \mathrm{e}^{-k\sb{vel}\ell} \simeq \Lambda (k\sb{vel}\ell)^2\ .
\ee
An approximate solution to this equation (with an accuracy better than $10\%$ for $\Lambda \ge 10$) is given   by \Eq{XB}. One can think that for the scales less than $k\sb{vel}$, there is the cumulative effect and hydrodynamic-like behavior. For the scales larger than $k\sb{vel}$ the vortices does not seem to feel each other much, and the separate vortex line behavior is important,  hence the Kelvin waves. Corrections to this simplest viewpoint will be discussed below.

\subsection{\label{sss:}Estimate  of the blending function}
Having at hand an estimate of the velocity filed~\eq{dv-int2} we can use \Eq{eq2} to find the blending function $g(k\ell)$.

\paragraph{The first step} is to obtain the kinetic energy density $ E_{11}(k)$ (per unit length), produced by one sin-distorted (with wave-vector $k$) vortex line:
\bse\label{est1}\bea\label{est1A}
 E_{11}(k) &=& \frac{\rho_0 }2\int\!\!\!\!\int \< \big|v_k (R,z)\big|^2\> \, dx dy~~~~~~~~ \\ %%
 \label{est1B}%%
 \approx \rho_0 \,  \frac{(A\, \kappa\,  \Lambda\,  k)^2}{8\pi}\!\!\!\!\!\!&&\!\!\!\!\!\!\int_0^\infty\! \frac{(1+ \pi \rho/2)\ e^{-2 \rho}}{(2+\Lambda\rho^2)^2}\, \rho \, d \rho \\ %%
 \label{est1C}%%
 \approx\ \rho_0 \,  \frac{\Lambda}{8\pi} \!\!\!\!\!\!&&\!\!\!\!\!\! \big(A\, \kappa\, k \big)^2\,, \quad \mbox{for \ }\Lambda\gg 1\ .~~ %%
\eea\ese
Here the fluid density $\rho_0$ should be distinguished from the dimensionless radius $\rho=k  R$, $R=\sqrt{x^2+y^2}$.  In \Eq{est1A} we should substitute  $v_k (R,z)= v_k (R)\cos(kz+\phi)$ from \Eq{dv-int2} integrate over $(x,y)$-plane orthogonal to the mean vortex line directed along $z$, and average $\<\cos^2(kz+\phi)\>=\frac 12$ along $z$. Considering the  integral in dimensionless polar coordinates $\rho$ and $\varphi$ after the free integration over $\varphi$, we get \Eq{est1B}, in which integral can be analytically taken in the limit of large
$\Lambda$ with the result~\eq{est1C}. This  formula can also be obtained directly from the  Kelvin-wave Hamiltonian in the so-called local-induced approximation, see e.g. Eq.~(5b) in \cite{lnr}, where one expands $\sqrt{1+|d w  (z)/ dz |^2}$ over the line distortion $w(z)=A \sin(kz)$.%%
 \paragraph{The second step} is to find cross-kinetic energy density $E_{1j}(k,\ell_{1j})$ (per unite length), proportional to the product of the velocities produced by two sin-distorted vortex lines, separated by $\ell_{1j} $.
 \bse\label{est2}\bea
\nonumber \!\!\!\!\!\!&&\!\!\!\!\!\! E_{1j}(k,\ell_{1j}) = \rho_0\!\int_{-\infty} ^\infty\!\!\!\! dx\!\! \int_{-\infty}^{\infty}\!\!\!\!dy   \< \big|v_k (R_1,z)v_k (R_2,z)\big|\>~~~~\\ \label{est2B}
\!\!\!\!\!\!&&   =  \frac {\rho_0}{2} \Big(\frac{\kappa A \Lambda k}{\pi}\Big)^2  \int\limits _0^\infty \d\~x
\int\limits_{-\infty}^{\~\ell_{1j}/2}\d\~y \, F(\rho_1) F(\rho_2)\ e^{-\rho\Sb{+}},\ ~~~ \\
\!\!\!\!\!\!&&\ \  F(\rho) = \frac{\sqrt{1 +\pi\rho/2  }}{2 +\Lambda \rho^2} \,, \quad \rho\Sb+=\rho_1+\rho_2\ .
 \eea
 Here $\~x=k\, x$,\ \ $\~ y= k\, y$,\ \  $ \~\ell_{1j}=k \ell_{1j}$,\   $\rho_1=k \sqrt{x^2+y^2}$\ \ and $\rho_2=k \sqrt{x^2+(y-\ell_{1j})^2}$. The change in the limits of integration is due to the (average) symmetry of the integrand relative to the plane $y=\ell_{1j}/2$. The main contribution to the integral comes from the region with $\~x\ \lesssim \~\ell_{1j}$, which allows to simplify the integrand replacing $ \rho_2  \Rightarrow \~\ell_{1j}-\~y+ \~x^2/ (2\,\ell_{1j})$. This clarifies the leading dependence  $E_{1j}(k,\ell_{1j})$ on $\ell_{1j}$ as%%%
 \be\label{est2E}
 E_{1j}(k,\ell_{1j})\propto \exp(- k\, \ell_{1j})\ .
 \ee%%
 Algebraic improvement of this estimate cane be obtained by neglecting the $(x,y)$-dependence of $F(\rho_2)$ in \Eq{est2B}. This procedure can be justified asymptotically for $k\ell_{1j}\gg 1$. The resulting dependence $E_{1j}(k,\ell_{1j}) \propto F(k \ell_{1j})\exp(- k\, \ell_{1j})$ approximates the integral \Eq{est2B} with an accuracy of 10\% for $k \,\ell_{1j}>2$. For smaller $k\ell_{1j}$ one has to account for $\ell_{1j}$ dependence of the remained integral $\int_{-\infty}^{\ell_{1j}/2} \dots d y \simeq (a/\Lambda)\ln (k \ell_{1j}+ b)$ with $a$ and $b$ weakly dependent on $\Lambda$.  As a result we approximate (with accuracy about 25\%) the $k\ell_{1j}$  dependence of the energy $E_{1j}$ as follows:
 \bea
 \nonumber E_{1j}(k,\ell_{1j}) &\approx&   \rho_0 \frac{\Lambda }{8\pi}\, \big(A\, \kappa\, k \big)^2\,  a \ln (k \ell_{1j}+b ) \times \\ %%
  \label{est2D}%%
   && \frac {\sqrt{1+\pi k \ell_{1j}/2}}{2+\Lambda (k\ell_{1j})^2}\, \exp(-k \ell_{1j})\ .
 \eea
In typical experiments with superfluid $^3$He and $^4$He the value of $\Lambda$ varies from 10 to 30. Therefore we present  here parameters  $a\approx 1.37$, $b\approx 1.25$ for $\Lambda=10$ and $a\approx 2.37$, $b\approx 1.17$
for $\Lambda=30$.
 \ese

  \paragraph{The third step} is to  find   the relative total cross-energy of all $j\ne 1$ pairs
  \BSE{est4}\BEA{est4A} R(k\ell)&=&\sum_{j\ne 1} \frac{E_{1j}(k,\ell_{1j})}{E_{11}(k)} %%
  \\ \label{est4B}%%
  &\simeq& \frac{4\pi}{(k\ell)^2} \int _{\ell_0}^\infty d \ell_{1j}\, \ell_{1j}\, \frac{E_{1j}(k,\ell_{1j})}{E_{11}(k)}\,,\eea%%
  \ese%%
  which is estimated in \Eq{est4B} in continuous approximation assuming that  $j$-lines are randomly distributed around line $i=1$  with mean density $1/\ell^2$ for $\ell_{1j}$ exceeding,   value $\ell$. Generally speaking, the integral~\eq{est4B} should contain probability function of vortex separations $\C P(\ell_{ij})$. Having no reasonable model for it, we choose a simple step function $\C P(x)=0$ for $x>1$.

 Blending function $g(k\ell)$, defined by \Eq{eq2}, is related to the ratio~\eq{est4} as follows:
 \BE{est5A} g(k\ell)= R(k\ell)\Big / [1+R(k\ell)]\,,
 \ee%%
 Taking  oversimplified representation~\eq{est2E} of integral~\eq{est2B} one computes $g_0(k\ell)$, see \Eq{est5B}:%%
 \begin{equation}%%
   g_0(k\ell) =  \Big [ 1+ \frac{(k\ell)^2\exp(  k\ell )}{4\pi (1+  k\ell)}\Big ]^{-1}\ .
 \end{equation}%%
 Comparing $\L$-independent function $g_0(k\ell)$ with  $g\sb{num}(k\ell)$, which depends also on $\L$, we have improved its analytical representation by introducing $\L$-dependent rescaling of the argument, see \Eq{est5C}:%%
 \begin{equation}%%
   g(k\ell)  =  g_0[0.32\,\ln (\L+7.5)\ k\ell]\ .
 \end{equation}%%
 The resulting function  $g(k\ell)$  gives a very reasonable approximation to the results of ``exact" numerical calculation of $g\sb{num}(k\ell)$.  Therefore in practical calculations we will use analytical form~\eq{est5} of the blending function  $g(k\ell)$.

 %%%%%%%%%%%%%%%%%%%%%%%%%%%%%%%%%%%%%%%%%%%%%%%%%%%%%%%%%%%%%%%%%%%%%%%%%%%%%%
  %%%%%%%%%%%%%%%%%%%%%%%%%%%%%%%%%%%%%%%%%%%%%%%%%%%%%%%%%%%%%%%%%%%%%%%%%%%%%%
   %%%%%%%%%%%%%%%%%%%%%%%%%%%%%%%%%%%%%%%%%%%%%%%%%%%%%%%%%%%%%%%%%%%%%%%%%%%%%%

\end{document}